\documentclass{iau} 
\usepackage{graphicx}

\title[IFU observations of the Tarantula] %% give here short title %%
{The Tarantula Nebula as a template for extragalactic star forming regions from VLT/MUSE and HST/STIS}

\author[Paul A. Crowther et al.]   %% give here short author list %%
{Paul A. Crowther$^1$, Saida M. Caballero-Nieves$^{1, 2}$, Norberto Castro$^3$ \and Christopher J. 
Evans$^{4}$}

\affiliation{$^1$Department of Physics \& Astronomy, University of Sheffield, Hounsfield Road,\\ 
Sheffield, S3 7RH, UK \\ email: {\tt Paul.Crowther@sheffield.ac.uk} \\[\affilskip]
$^2$Physics \& Space Sciences, Florida Institute of Technology, 150 W. University Blvd, Melbourne, FL 
32901, USA\\
$^{3}$Department of Astronomy, University of Michigan, 1805 S.University,\\ Ann Arbor, MI 48109, USA\\
$^{4}$UK Astronomy Technology Centre, Royal Observatory Edinburgh, Blackford Hill,\\ Edinburgh, EH9 
3HJ, 
UK}

\pubyear{2017}
\volume{329}  %% insert here IAU Symposium No.
\jname{The lives and death-throes of massive stars}
\editors{J.J. Eldridge et al., ed.}
\begin{document}

\maketitle

\begin{abstract}
We present VLT/MUSE observations of NGC 2070, the dominant ionizing nebula of 30 Doradus in the LMC, 
plus HST/STIS spectroscopy of its central star cluster R136. Integral Field Spectroscopy (MUSE) and
pseudo IFS (STIS) together provides a complete census of all massive stars within the central 30$\times$30 
parsec$^2$ of the Tarantula. We discuss the integrated far-UV spectrum of 
R136, of particular interest for UV studies of young extragalactic star clusters. Strong He\,{\sc ii} 
$\lambda$1640 emission at very early ages (1--2 Myr) from very massive stars cannot be reproduced by 
current population synthesis models, even those incorporating binary evolution and very massive stars. A nebular
analysis of the integrated MUSE dataset implies an age of $\sim$4.5 Myr for NGC 2070. Wolf-Rayet features provide
alternative age diagnostics, with the primary contribution to the integrated Wolf-Rayet bumps arising from R140 rather 
than the more numerous H-rich WN stars in R136. Caution should be used when interpreting spatially extended observations
of extragalactic star-forming regions. 
\end{abstract}

\firstsection % if your document starts with a section,
\section{Introduction to the Tarantula Nebula}

Our interpretation of distant, unresolved star-forming regions relies heavily upon population synthesis models,
including Starburst99 (\cite[Leitherer et al. 2014]{Leitherer14}) and BPASS (\cite[Eldridge \& Stanway 2012]{Eldridge12}). However, such models
rely on a number of key assumptions, involving the initial mass function (IMF), rotation, star formation history.
In particular, the post-main sequence evolution of massive stars depends sensitively on internal mixing processes and
adopted mass-loss prescriptions. Locally, close binary evolution looks to play a major role (\cite[Sana et al. 2012]{Sana12})
and there is some evidence 
that the upper IMF extends well beyond the usual $M_{\rm max} \sim 100 M_{\odot}$ (\cite[Crowther et al. 2010]{Crowther10}).

Therefore, it is important to benchmark population synthesis models against empirical results from nearby, resolved
star-forming regions. Within the Local Group of galaxies, the LMC's Tarantula Nebula is the
closest analogue to the intensive star-forming clumps of high-redshift galaxies (\cite[Jones et al. 2010]{Jones10}).
\cite[Kennicutt (1984)]{Kennicutt84} compares the nebular properties of nearby extragalactic HII regions, including the Tarantula
(alias 30 Doradus). Although the Tarantula Nebula extends over several hundred parsecs, the central ionizing
region, NGC\,2070, spans 40 pc, with the massive, dense cluster R136\footnote{Strictly, the star cluster is R136a, with R136b and R136c representing 
individual very massive stars, but R136 is generally used to describe the cluster} at its core (see Table 1 of \cite[Walborn 1991]{Walborn91}).

\begin{figure}
\begin{center}
\includegraphics[width=90mm]{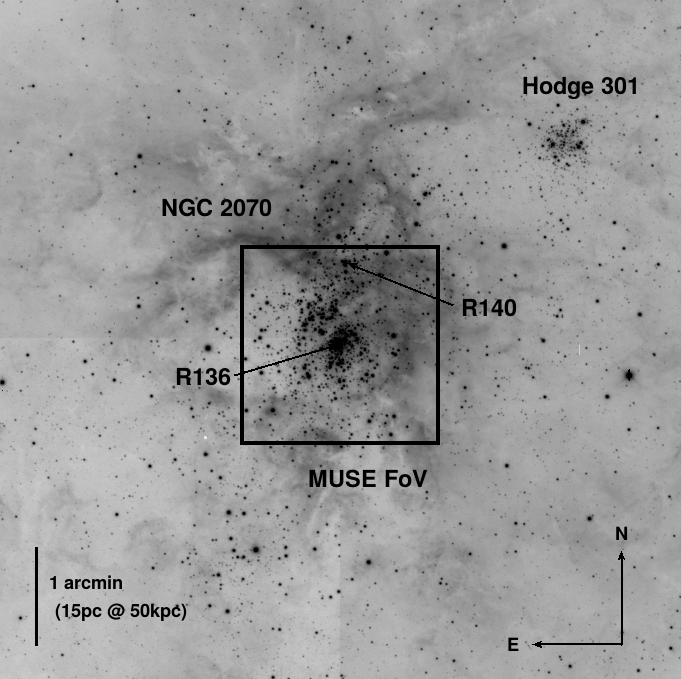} 
\caption{VLT/FORS2 R-band image of the central 6$\times$6 arcmin$^{2}$ of the Tarantula Nebula, corresponding to 100$\times$100 parsec
at the 50kpc LMC distance, indicating NGC 2070, R136, R140, Hodge~301 and the VLT/MUSE field of view.}
\label{30Dor}
\end{center}
\end{figure}

In this work, we exploit the high spatial resolution of HST to dissect the central 4$\times$4 arcsec$^{2}$ of the 
R136 star cluster  (\cite[Crowther et al. 2016]{Crowther16}) plus the new VLT large field-of-view integral 
field spectrograph MUSE which has observed the central  2$\times$2 arcmin$^{2}$ region of NGC 2070 (N.~Castro et al. in prep).
Figure~\ref{30Dor} shows a VLT/FORS2 R-band image of the central region of the Tarantula, including NGC~2070 and an older
star cluster (Hodge 301) 3$'$ to the NW.  Comparisons with population synthesis predictions are made to assess their validity for 
young starburst regions.

\begin{figure}
\begin{center}
  \includegraphics[width=70mm,angle=-90]{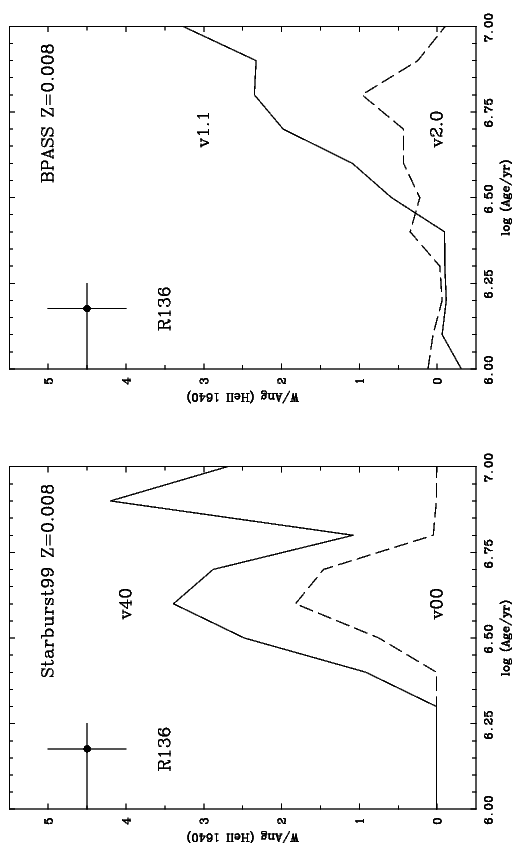}
  \caption{Comparison between observed He\,{\sc ii} $\lambda$1640 emission in R136 (\cite[Crowther et al. 2016]{Crowther16}) and
    Z=0.008 instantaneous
    burst models from (left): Starburst99 (\cite[Leitherer et al. 2014]{Leitherer14}) for non-rotating (v00, dashed) and rotating (v40, solid) 
Geneva single star models with $M_{\rm max} = 120 M_{\odot}$, (right):
    BPASS (\cite[Eldridge \& Stanway 2012]{Eldridge12},
    \cite[Stanway et al. 2016]{Stanway16}) close binary models for $M_{\rm max} = 100 M_{\odot}$ (v1.1, solid) and $M_{\rm max} = 300 M_{\odot}$ 
(v2.0, dashed).}
\label{HeII1640}
\end{center}
\end{figure}

\section{HST/STIS ultraviolet spectroscopy of R136}

R136 is the young massive star cluster at the core of the Tarantula Nebula. Establishing the stellar content of R136 has generally required high 
spatial spectroscopy from HST at UV and optical wavelengths or a ground-based 8m telescope with Adaptive Optics. \cite[Massey \& Hunter (1998)]{Massey98} have used 
HST/FOS to observe stars in the R136 region (few parsec), while \cite[Crowther et al. (2016)]{Crowther16} used HST/STIS to map the core of R136,
obtaining complete UV/optical spectroscopy of the central parsec. Spectroscopic analysis of the brightest 52 O stars reveals a cluster 
age of 1.5$^{+0.3}_{-0.7}$ Myr, confirming  this as one of the youngest massive clusters in the Local Group. 

The integrated UV spectrum of R136 exhibits strong stellar C\,{\sc iv}  $\lambda\lambda$1548-51, N\,{\sc v}  $\lambda\lambda$1238-42 P  Cygni 
profiles plus prominent 
He\,{\sc ii} $\lambda$1640 emission. Within unresolved high mass star clusters, He\,{\sc ii} $\lambda$1640 emission is usually believed to 
originate 
from 
classical 
He-burning  Wolf-Rayet (WR) stars which become prominent after several Myr. The presence of strong He\,{\sc ii} emission in the young R136 
indicates a 
different origin, namely the presence of very massive stars (VMS, $M_{\rm init} \geq 100 M_{\odot}$). Indeed, $\sim$10 VMS within the central parsec 
contribute one third of the integrated far-UV continuum, and dominate the observed He\,{\sc ii} emission. 

Standard population synthesis models fix the upper mass limit at $M_{\rm max} \sim 100 M_{\odot}$, so fail to produce significant He\,{\sc ii} 
$\lambda$1640 emission until $\sim$2--3 Myr after the initial burst, as illustrated in the left panel of Fig~\ref{HeII1640}, which compares
predictions from Starburst99 based on metal-poor (Z=0.008) Geneva models (\cite[Leitherer et al. 2014]{Leitherer14}) with the 
observed strong, broad He\,{\sc ii} emission in R136 (\cite[Crowther et al. 2016]{Crowther16}). These predictions are based on single, non-rotating 
(v00) or rotating (v40) 
models, so exclude the contribution of close binary evolution and VMS. Close binary evolution has been included into metal-poor Z=0.008 BPASS burst 
models (v1.1, \cite[Eldridge \& Stanway 2012]{Eldridge12}), which is presented on the right panel of Fig.~\ref{HeII1640} together with recent updates (v2.0, \cite[Stanway 
et al. 2016]{Stanway16}) in which very massive stars are incorporated. Weaker emission in the more recent models for ages of $\geq$3 Myr results 
from the use of different spectral libraries.

It is apparent that both single and binary population synthesis models currently fail to reproduce the observed strong He\,{\sc ii} $\lambda$1640 
in R136, 
indicating an incomplete treatment of rotation, binarity and mass-loss for the most massive stars in star clusters. In the nearby universe, a number 
of young massive star clusters also exhibit strong He\,{\sc ii} $\lambda$1640 emission (\cite[Wofford et al. 2014]{Wofford14}, \cite[Smith et al. 
2016]{Smith16}) while it is a prominent
(stellar and nebular) spectral feature in the integrated rest-frame UV light of high redshift Lyman Break galaxies (\cite[Shapley et al. 2003]{Shapley03}). VMS and 
classical WR stars will initially play a major role in stellar He\,{\sc ii} $\lambda$1640 emission, with close binary evolution subsequently 
dominant.

\begin{figure}
\begin{center}
\includegraphics[width=100mm]{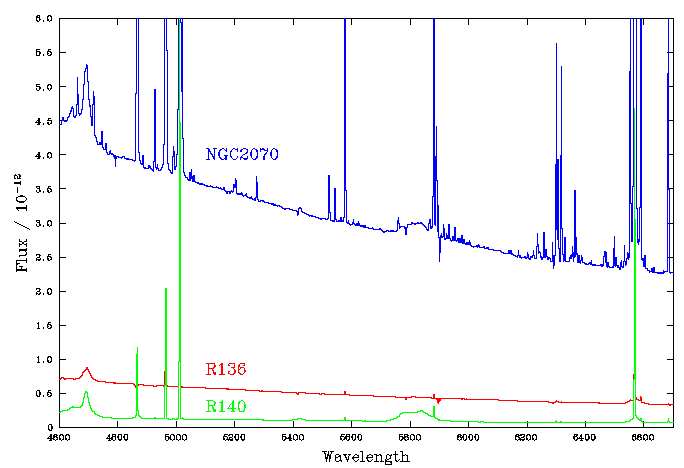} 
\caption{Integrated VLT/MUSE spectrum of NGC~2070 from N.~Castro et al. (in prep), highlighting strong
  blue (He\,{\sc ii} $\lambda$4686, C\,{\sc iii} $\lambda\lambda$4647-51) and yellow (C\,{\sc iv} $\lambda\lambda$5801-12) 
Wolf-Rayet features, including the individual contributions from R136 and R140.}
\label{MUSE_clusters}
\end{center}
\end{figure}

\section{VLT/MUSE integral field spectroscopy of NGC 2070}

Individual massive stars within NGC 2070 -- the dominant HII region within the Tarantula Nebula -- have been 
extensively  investigated from the ground (\cite[Melnick 1985]{Melnick85}, \cite[Walborn \& Blades 1997]{Walborn97}, \cite[Bosch et al. 1999]{Bosch99}, \cite[Evans et al. 2011]{Evans11}), but the advent of large 
field-of-view integral field spectrographs such as MUSE permit the integrated properties of nearby giant HII regions to be investigated.
We have obtained four pointings of VLT/MUSE which cover the central region of NGC~2070 (Fig.~\ref{30Dor}), and provide medium resolution optical
spectroscopic observations (R=3,000, $\lambda$=4600--9360\AA) of the 2$\times$2 arcmin$^{2}$ region, sampled at 0.3 arcsec\,spaxel$^{-1}$. 

Nebular and stellar kinematics of NGC~2070 are discussed in N.~Castro et al. (in prep), while the individual stellar content will be discussed 
elsewhere. 
Here we focus upon the integrated stellar and nebular properties of the VLT/MUSE dataset, which samples a 30$\times$30 parsec$^{2}$ region of 
NGC~2070, including R136. At a distance of 10 Mpc, this region would subtend only 0.6 arcsec, typical of long-slit spectroscopy of extragalactic
HII regions from large ground-based telescopes. The integrated VLT/MUSE spectrum of NGC~2070 is presented in Fig.~\ref{MUSE_clusters}, revealing 
prominent nebular emission features. An  interstellar extinction of E(B-V) = 0.38 follows from the observed H$\alpha$/H$\beta$ flux ratio, in good 
agreement with \cite[Pellegrini et al. (2010)]{Pellegrini10},  while the measured H$\alpha$ flux is 60\% of that measured by \cite[Kennicutt et al. 
(1995)]{Kennicutt95} for a 3$'$ radius centred upon R136, such that the inferred H$\alpha$ luminosity, $\log L$(H$\alpha$) = 39.1 erg\,s$^{-1}$, 
corresponds to 10\% of  the total emission from the Tarantula Nebula (\cite[Kennicutt 1984]{Kennicutt84}). 

Commonly used diagnostics to estimate the age of young extragalactic star forming regions are nebular Balmer emission and the presence of WR
features in the integrated light.  We infer an age of $\sim$4.5 Myr from $\log W_{\lambda}$ (H$\alpha$)/\AA\ = 2.85 for NGC~2070, even though it 
is 
clear 
that the individual ages of massive stars within NGC~2070 span a broad range (e.g. \cite[Selman et al. 1999]{Selman99}, F.~Schneider et al. in prep)
with  massive star  formation still ongoing 
in NGC~2070 (\cite[Walborn et al. 2013]{Walborn13}) 
plus Hodge 301 to the  NW of NGC~2070 indicating a burst of star formation 15--25 Myr ago. Recalling 
Fig.~\ref{MUSE_clusters}, both the blue (He\,{\sc ii} $\lambda$4686, C\,{\sc iii} $\lambda\lambda$4647-51) and yellow (C\,{\sc iv}  
$\lambda\lambda$5801-12) WR features are prominent in the integrated VLT/MUSE dataset, with $W_{\lambda}$(blue) $\sim$12\AA, and 
$W_{\lambda}$(yellow) $\sim$9\AA.
As with He\,{\sc ii} $\lambda$1640 in ultraviolet spectroscopy, it is usually assumed that WR bumps arise from classical He-burning stars. Indeed, 
for a 
burst
age of 4--5 Myr, Starburst99 models at Z=0.008 predict $W_{\lambda}$(blue)$\sim$15\AA\ (5\AA) and $W_{\lambda}$(yellow)$\sim$6\AA\ (3\AA) from v40 
rotating  
(v00 non-rotating) models. 

Therefore, at face value it appears that a single 4--5 Myr burst is reasonably consistent with the observed nebular and stellar line features 
in NGC~2070. However, integral field spectroscopy also permits us to consider the origin of these WR features within NGC~2070. 
Fig.~\ref{MUSE_clusters} also displays the integrated spectrum of R136, which exhibits weak He\,{\sc ii} $\lambda$4686 emission from the VMS in 
this 
young $\sim$1.5 Myr cluster, contributing only 15\% of the blue NGC~2070 WR feature, and crucially negligible emission at C\,{\sc iv} 
$\lambda\lambda$5801--12 since it is too young to host any WC stars. Indeed, the dominant source of the yellow WR feature in NGC~2070 is R140, a 
relatively modest group of stars including two classical WN stars and a WC star (indicated in Fig.~\ref{30Dor}, while R140 also contributes 25\% of 
the integrated blue  WR feature. The higher line luminosities of classical WR stars (in R140) with respect to H-rich WN stars (in R136) 
compensate for their reduced population.

Therefore caution should be used when interpreting spatially extended regions in extragalactic star forming regions. In the absence of spatially 
resolved spectroscopy, one would anticipate that the integrated stellar and nebular properties of NGC~2070 are dominated by the R136 region, whereas 
the nebular-derived age represents a composite of the young R136 cluster and older OB stars, while the stellar-derived age is biased towards WR stars 
possessing the highest line luminosities (R140), rather than the those within the dominant ionizing cluster (R136). By way of example, the dominant
cluster in NGC~3125-A has an age of 1--2 Myr (\cite[Wofford et al. 2014]{Wofford14}) while a burst age of 4 Myr is inferred from its
associated star-forming region from nebular and WR diagnostics (\cite[Hadfield \& Crowther 2006]{Hadfield06}).

%\section{Summary}

%\begin{figure}
%\includegraphics[width=120mm]{r136_slits.png} 
%\caption{Overlay of HST/STIS slits on HST/WFC3 image of R136 and its immediate surroundings,
%providing pseudo integral field spectroscopy of the star cluster (Crowther  et al. 2016).}
%\label{r136_slits}
%\end{figure}

%\begin{figure}
%\includegraphics[width=120mm]{r136_cluster.png} 
%\caption{Integrated spectrum of R136 from HST/STIS spectroscopy, showing the breakdown between very 
%massive stars ($\geq$ 100 $M_{\odot}$) and other stars (Crowther et al. 2016).}
%\label{r136_cluster}
%\end{figure}

\end{document}